\documentclass[11pt]{article}\textwidth 6.5in\textheight 9in
\usepackage{amssymb}\usepackage[colorlinks]{hyperref}\usepackage{color}
\usepackage{stmaryrd}\usepackage{mathrsfs}
\usepackage{graphicx}
\usepackage{amsmath}
\usepackage{caption}

\usepackage{braket}

\begin{document}
	\setlength{\captionmargin}{27pt}
	\newcommand\hreff[1]{\href {http://#1} {\small http://#1}}
	\newcommand\trm[1]{{\bf\em #1}} \newcommand\emm[1]{{\ensuremath{#1}}}
	\newcommand\prf{\paragraph{Proof.}}\newcommand\qed{\hfill\emm\blacksquare}
	
	\newtheorem{thr}{Theorem} 
	\newtheorem{lmm}{Lemma}
	\newtheorem{cor}{Corollary}
	\newtheorem{con}{Conjecture} 
	\newtheorem{prp}{Proposition}
	
	\newtheorem{blk}{Block}
	\newtheorem{dff}{Definition}
	\newtheorem{asm}{Assumption}
	\newtheorem{rmk}{Remark}
	\newtheorem{clm}{Claim}
	\newtheorem{example}{Example}
	
	\newcommand{\Z}{\mathbb{Z}}
	\newcommand{\ab}{a\!b}
	\newcommand{\yx}{y\!x}
	\newcommand{\yux}{y\!\underline{x}}
	\newcommand{\Tr}{\mathrm{Tr}}
	\newcommand\floor[1]{{\lfloor#1\rfloor}}\newcommand\ceil[1]{{\lceil#1\rceil}}

	\newcommand{\lea}{<^+}
	\newcommand{\gea}{>^+}
	\newcommand{\eqa}{=^+}

	\newcommand{\lel}{<^{\log}}
	\newcommand{\gel}{>^{\log}}
	\newcommand{\eql}{=^{\log}}
	
	\newcommand{\lem}{\stackrel{\ast}{<}}
	\newcommand{\gem}{\stackrel{\ast}{>}}
	\newcommand{\eqm}{\stackrel{\ast}{=}}
	
	\newcommand\edf{{\,\stackrel{\mbox{\tiny def}}=\,}}
	\newcommand\edl{{\,\stackrel{\mbox{\tiny def}}\leq\,}}
	\newcommand\then{\Rightarrow}
	
	\newcommand\km{{\mathbf {km}}}\renewcommand\t{{\mathbf {t}}}
	\newcommand\KM{{\mathbf {KM}}}\newcommand\m{{\mathbf {m}}}
	\newcommand\md{{\mathbf {m}_{\mathbf{d}}}}\newcommand\mT{{\mathbf {m}_{\mathbf{T}}}}
	\newcommand\K{{\mathbf K}} 
	\newcommand\I{{\mathbf I}}
	\newcommand\Iu{\overline{\mathbf I}}
	\newcommand\Il{\underline{\mathbf I}}
	\newcommand\Kd{{\mathbf{Kd}}} \newcommand\KT{{\mathbf{KT}}} 
	\renewcommand\d{{\mathbf d}} 
	\newcommand\du{\overline{\mathbf d}}
	\newcommand\dl{\underline{\mathbf d}}
	\newcommand\D{{\mathbf D}}
	
	\newcommand\w{{\mathbf w}}
	\newcommand\Ks{\Lambda} \newcommand\q{{\mathbf q}}
	\newcommand\E{{\mathbf E}} \newcommand\St{{\mathbf S}}
	\newcommand\M{{\mathbf M}}\newcommand\Q{{\mathbf Q}}
	\newcommand\ch{{\mathcal H}} \renewcommand\l{\tau}
	\newcommand\tb{{\mathbf t}} \renewcommand\L{{\mathbf L}}
	\newcommand\bb{{\mathbf {bb}}}\newcommand\Km{{\mathbf {Km}}}
	\renewcommand\q{{\mathbf q}}\newcommand\J{{\mathbf J}}
	\newcommand\z{\mathbf{z}}
	
	\newcommand\B{\mathbf{bb}}\newcommand\f{\mathbf{f}}
	\newcommand\hd{\mathbf{0'}} \newcommand\T{{\mathbf T}}
	\newcommand\R{\mathbb{R}}\renewcommand\Q{\mathbb{Q}}
	\newcommand\N{\mathbb{N}}\newcommand\BT{\Sigma}
	\newcommand\FS{\BT^*}\newcommand\IS{\BT^\infty}
	\newcommand\FIS{\BT^{*\infty}}\newcommand\C{\mathcal{L}}
	\renewcommand\S{\mathcal{C}}\newcommand\ST{\mathcal{S}}
	\newcommand\UM{\nu_0}\newcommand\EN{\mathcal{W}}
	\newcommand\QC{\mathbf{R}}
	\newcommand{\supp}{\mathrm{Supp}}
	
	\newcommand\lenum{\lbrack\!\lbrack}
	\newcommand\renum{\rbrack\!\rbrack}
	
	\newcommand\uhr{\upharpoonright}
	\renewcommand\qed{\hfill\emm\square}
	\renewcommand\Tr{\mathrm{Tr}}
	\renewcommand\H{\mathbf{H}}
	\renewcommand\m{\mathbf{m}}
	
	\newcommand{\bmu}{\boldsymbol{\mu}}
	
	\renewcommand\i{{\mathbf i}}
	\newcommand\Hl{{\mathbf H}}
	
	\newcommand\ml{\underline{\mathbf m}}
	\newcommand\mup{\overline{\mathbf m}}
	\title{\vspace*{-3pc}  On the Algorithmic Content of Quantum Measurements}
	
	\author {Samuel Epstein\footnote{
		JP Theory Group,
		samepst@jptheorygroup.org}}
	
	\maketitle
	\begin{abstract}
		 We show that given a quantum measurement, for an overwhelming majority of pure states, no meaningful information is produced. This is independent of the number of outcomes of the quantum measurement. Due to conservation inequalities, such random noise cannot be processed into coherent data.
	\end{abstract}
\section{Introduction}

Quantum information theory studies the limits of communicating through quantum channels. In \cite{Holevo73}, the Holevo bound was proven, providing an upper bound on the amount of classical information shared between two parties that can prepare and measure mixed states. The Holevo bound states that only $n$ bits of classical information can be accessed from $n$ qubits.
The main result of this paper shows the limitations of the algorithmic content of measurements of pure quantum states. Given a measurement apparatus $E$, there is only a tiny fraction of quantum pure states on which $E$’s application produces coherent information. This is independent of the number of measurement outcomes of $E$.

In this paper we introduce a novel definition of algorithmic information between probabilities. This definition uses the information of two strings $x$ and $y$, $\i(x:y)$, defined in Section \ref{sec:conv}.\\
 
\noindent\textbf{Definition.}\textit{ The amount of algorithmic information between two probabilities $p$ and $q$ over $\FS$ is 
$\i(p:q)=\log\sum_{x,y\in\FS}2^{\i(x\,{:}\,y)}p(x)q(y)$.}\\

If a probability $p$ has low self information $\i(p:p)$ then it has no meaningful information. Generally speaking, such probabilities either have high measure for simple strings and/or very low measure over complex strings. As shown in the appendix, the information between two probabilities is conserved over randomized transformations. Therefore there are no means to increase the self-information of a probability.

In quantum mechanics, given a quantum state $\ket{\psi}$, a measurement, or POVM, $E$ produces a probability measure $E\ket{\psi}$ over strings. This probability represents the classical information produced from the measurement. The exact structure of POVMs is described in Section \ref{sec:measure}. The main theorem of this paper states that given a measurement $E$, for an overwhelming majority of quantum states $\ket{\psi}$, the probability produced will have no meaningful information, i.e. $\i(E\ket{\psi}:E\ket{\psi})$ is negligible. \\

\noindent\textbf{Theorem}.  \textit{Let $\Lambda$ be the uniform distribution on the unit sphere of an $n$ qubit space. Relativized to POVM $E$, $\int 2^{\i(E\ket{\psi}:E\ket{\psi})}d\Lambda = O(1)$.}\\

\section{Related Work}
For information about the history and foundation of algorithmic information theory, we refer readers to the textbooks \cite{DowneyHi10} and \cite{LiVi08}. There are several definitions that model the algorithmic content of a quantum state. In \cite{BerthiaumeVaLa01}, the complexity of a quantum state is equal to the size of the smallest quantum Turing machine that can approximate the state to a given fidelity. In \cite{MoraBe05}, the algorithmic complexity of a quantum state is equal to the minimal length of an encoding of the preparation of the state through quantum gates. In \cite{Gacs01}, the algorithmic entropy of a quantum state is measured by the negative logarithmic of the state multiplied by a universal lower computable semi-density matrix. In \cite{Vitanyi00}, the entropy of a pure quantum state is equal to the classical complexity of an elementary approximating state plus the negative logarithm of their fidelity. A quantum version of Brudno's theorem was proven in \cite{Benatti06}. Randomness for infinite quantum spin chains, called quantum Martin L\"{o}f random sequences, was introduced in \cite{NiesSc19}. An infinite version of algorithmic entropy can be found at \cite{Benatti14}. 
	\section{Conventions}
	\label{sec:conv}
We use $\FS$ to denote finite strings. The length of a string $x\in\FS$ is $\|x\|$. For positive real function $f$, $\lea f$, $\gea f$, and $\eqa f$ is used to represent $< f+O(1)$, $>f+O(1)$, and $=f\pm O(1)$.  In addition ${\lem}f$, ${\gem}f$ denote $<f/O(1)$, $>f/O(1)$. The terms ${\eqm}f$  denotes ${\lem}f$ and ${\gem}f$. The encoding of $x\in\FS$ is $\langle x\rangle=1^{\|x\|}0x$. 

For strings $x,y\in\FS$, the output of algorithm $T$ on input $x$ and auxiliary input $y$ is denoted $T_y(x)$. An algorithm $T$ is prefix free if for strings $x,y,s\in\FS$, $\neq\emptyset$, if $T_y(x)$ halts then $T_y(xs)$ does not halt. There exists a universal prefix free algorithm $U$, where for all prefix-free algorithms $T$, there exists a $t\in\FS$, where for all $x,y\in\FS$, $U_y(tx)=T_y(x)$. This $U$ is used to define Kolmogorov complexity, with $\K(x/y)=\min\{\|p\|:U_y(x)=p\}$. The universal probability of $x\in\FS$, conditional to $y\in\FS$, is $\m(x/y)=\sum\{2^{-\|p\|}:U_y(p)=x\}$. The algorithmic information between two strings is $\i(x:y)=\K(x)+\K(y)-\K(x,y)$. The expression ``relativized to an elementary object'' seen in theorems is equivalent to saying there is an encoding of the elementary object on an auxiliary tape of the universal algorithm $U$. 

We use $\mathcal{H}_n$ to denote a Hilbert space with $n$ dimensions, spanned by bases $\ket{\beta_1},\dots,\ket{\beta_n}$. A qubit is a unit vector in the Hilbert space $\mathcal{G}=\mathcal{H}_2$, spanned by vectors $\ket{0}$, $\ket{1}$. To model $n$ qubits, we use a unit vector in $\mathcal{H}_{2^n}$, spanned by basis vectors $\ket{x}$, where $x$ is a string of size $n$. 

A pure quantum state $\ket{\psi}$ of length $n$ is a unit vector in $\mathcal{H}_{2^n}$. Its corresponding element in the dual space is denoted by $\bra{\psi}$. The conjugate transpose of a matrix $A$ is $A^*$. The tensor product of two matrices $A$ and $B$ is $A \otimes B$. $\Tr$ is used to denote the trace of a matrix, and for Hilbert space $\mathcal{H}_X\otimes\mathcal{H}_Y$, the partial trace with respect to $\mathcal{H}_Y$ is $\Tr_Y$.  

For positive semi-definite matrices $A$ and $B$, we say $B \preceq  A$, if $A-B$ is positive semi-definite. For functions $f$ whose range are Hermitian matrices, we use ${\lem}f$ and ${\gem f}$ to denote $\preceq f/O(1)$ and $\succeq f/O(1)$. We use ${\eqm f}$ to denote ${\lem}f$ and ${\gem f}$. 

Density matrices are used to represent mixed states, and are self-adjoint, positive definite matrices with trace equal to 1. Semi-density matrices are density matrices except they may have a trace in [0,1]. 

Pure and mixed quantum states are elementary if their values are complex numbers with rational coefficients, and thus they can be represented with finite strings. Thus elementary quantum states $\ket{\phi}$ and $\rho$ can be encoded as strings, $\langle\ket{\phi}\rangle$ and $\langle\rho\rangle$, and assigned Kolmogorov complexities $\K(\ket{\phi})$, $\K(\rho)$ and algorithmic probabilities $\m(\ket{\phi})$ and $\m(\rho)$. They are equal to the complexity (and algorithmic probability) of the strings that encodes the states. 

More generally, a complex matrix $A$ is elementary if its entries are complex numbers with rational coefficients and can be encoded as $\langle A\rangle$, and has a Kolmogorov complexity $\K(A)$ and algorithmic probability $\m(A)$.

In \cite{Gacs01}, a universal lower computable semi-density matrix, $\bmu$ was introduced. It can be defined (up to a multiplicative constant) by
$\bmu = \sum_{\textrm{ elementary }\ket{\phi}}\m(\ket{\phi}/n)\ket{\phi}\bra{\phi},$
where the summation is over all $n$ qubit elementary pure quantum states.

We say a semi-density matrix $\rho$ is lower computable if there a program $p\in\FS$ such that when given to the universal Turing machine $U$, outputs, with or without halting, a finite or infinite sequence of elementary matrices $\rho_i$ such that $\rho_i\preceq \rho_{i+1}$ and $\lim_{i\rightarrow\infty}\rho_i=\rho$. If $U$ reads $\leq \|p\|$ bits on the input tape, then we say $p$ lower computes $\rho$. From \cite{Gacs01} Theorem 2, if $q$ lower computes $\rho$, then $\m(q/n)\rho\lem\bmu$. 

We say a semi-density matrix $\rho\otimes \sigma$ is upper computable if there a program $p\in\FS$ such that when given to the universal Turing machine $U$, outputs, with or without halting, a finite or infinite sequence of elementary matrices $\rho_i\otimes \sigma_i$ such that $\rho_{i+1}\otimes\sigma_{i+1}\preceq \rho_{i}\otimes\sigma_i$ and $\lim_{i\rightarrow\infty}\rho_i\otimes\sigma_i=\rho\otimes\sigma$. If $U$ reads $\leq \|p\|$ bits on the input tape, then we say $p$ upper computes $\rho\otimes\sigma$. 
The upper probability of an upper computable mixed state $A\otimes B$ is defined by $\mup(A\otimes B/x)=\sum\{\m(q/x):q\textrm{ upper computes }A\otimes B\}$.

\section{Information}
 Let $\mathcal{C}_{C\otimes D}$ be the set of all upper computable matrices (tests) of the form $A\otimes B$, where $\Tr(A\otimes B)(C\otimes D)\leq 1$. Let $\mathfrak{C}_{C\otimes D}=\sum_{A\otimes B\in \mathcal{C}_{C\otimes D}}\mup(A\otimes B/n)(A\otimes B)$ be an aggregation of upper computable $C\otimes D$ tests of the form $A\otimes B$, weighted by their upper  probability.
The information between semi-density matrices $A$ and $B$ is $\I(A:B)= \log \Tr \mathfrak{C}_{\bmu\otimes \bmu}(A\otimes B)$.

\begin{prp} $\I(2^{-n}:2^{-n})=O(1) $.
	\label{prp:infouppperlower}
\end{prp}
\prf 
$1\geq \Tr\mathfrak{C}_{\bmu\otimes\bmu}(\bmu\otimes\bmu)\gem \Tr \mathfrak{C}_{\bmu\otimes\bmu}(2^{-n}I\otimes 2^{-n}I)\gem 2^{\I(2^{-n}I:2^{-n}I)}$.
	
\qed

\begin{lmm}
	\label{lmm:selfinfo}
	Let $\Lambda$ be the uniform distribution on the unit sphere of an $n$ qubit space.\\ $\int 2^{\I(\ket{\psi}\,{:}\,\ket{\psi})}d\Lambda=O(1)$.
\end{lmm}

\prf Using \cite{Gacs01} Section 5 and \cite{BerthiaumeVaLa01} Section 6.3,  we have that $\int \,{\ket{\psi}}{\bra{\psi}}\otimes {\ket{\psi}\bra{\psi}}\,d\Lambda = \int \,{\ket{\psi\psi}}{\bra{\psi\psi}}\,d\Lambda$ $= {2^n+1\choose 2}^{-1}P$, where $P$ is the projection onto the space of pure states ${\ket{\psi\psi}}$. Using Proposition \ref{prp:infouppperlower},

\begin{align*}
	\int 2^{\I(\ket{\psi}\,{:}\,\ket{\psi})}d\Lambda &= \int \Tr\mathfrak{C}_{\bmu\otimes\bmu}\,{\ket{\psi}}{\bra{\psi}}\otimes {\ket{\psi}\bra{\psi}}\,d\Lambda=\Tr\mathfrak{C}_{\bmu\otimes\bmu} \int{\ket{\psi}}{\bra{\psi}}\otimes {\ket{\psi}\bra{\psi}}\,d\Lambda\\ 
	&= \Tr\mathfrak{C}_{\bmu\otimes\bmu}{2^n+1\choose 2}^{-1}P\lem \Tr\mathfrak{C}_{\bmu\otimes\bmu}2^{-2n}I\eqm 2^{\I(2^{-n}I:2^{-n}I)}=O(1).
	\end{align*}
\qed

\section{Measurements}
\label{sec:measure}
A POVM $E$ is a finite set of positive definite matrices $\{E_k\}$ such that $\sum_k E_k = I$. For a given density matrix $\sigma$, a POVM $E$ induces a probability measure over strings, where $E\sigma(k) = \Tr E_k\sigma$. This can be seen as the probability of seeing measurement $k$ given quantum state $\sigma$ and measurement $E$. An elementary POVM has each $E_k$ being elementary. We introduce a novel definition to algorithmic information theory, the amount of algorithmic mutual information between two probabilities. 
\begin{dff}[Information, Probabilities]$ $\\
	For probabilities $p$ and $q$ over $\FS$, $ \i(p:q)=\log\sum_{x,y\in\FS}2^{\i(x\,{:}\,y)}p(x)q(y)$.
\end{dff}
 Theorem \ref{the:classconsinfo} in the appendix proves conservation of information of probabilities transformed by random channels. A channel $f:\FS\times\FS\rightarrow\R$, such that $f(\cdot|x)$ is a probability for all $x\in\FS$, transforms a probability $p$ by $fp(x)=\sum_yf(x|y)p(y)$. Conservation occurs over $f$, $\i(fp:q)\lea \i(p:q)$. 

\begin{lmm}
	\label{lmm:meas}
	Relativized to POVM $E$,
	$\i(E\ket{\psi}{:}E\ket{\psi})\lea \I(\ket{\psi}{:}\ket{\psi})$.
\end{lmm}
\prf 
Since  $z(k)=\Tr\bmu E_k$ is lower semi-computable and $\sum_kz(k)< 1$, $\m(k/n) \gem \Tr\bmu E_k$, and so $1 > 2^{\K(k/n)-O(1)}\Tr\bmu E_k$. So $\nu_{i,j}=2^{\K(i/n)+\K(j/n)-O(1)}(E_i\otimes E_j)\in\mathcal{C}_{\bmu\otimes\bmu}$, with $\mup(\nu_{i,j}/n)\gem\m(i,j/n)$. 
\begin{align*}
\I(\ket{\psi}{:}\ket{\psi})&=\log \sum_{A\otimes B\in\mathcal{C}_{\bmu\otimes\bmu}}\mup(A\otimes B/n)(A\otimes B)(\ket{\psi}\bra{\psi}\ket{\psi}\bra{\psi})\\
&\gea \log\Tr \sum_{ij}\nu_{i,j}\mup(\nu_{i,j}/n)(\ket{\psi}\bra{\psi}\ket{\psi}\bra{\psi})\\
&\gea \log \sum 2^{\K(i/n)+\K(j/n)}\m(i,j/n)E\ket{\psi}(i)E\ket{\psi}((j)\\
&\gea \i(E\ket{\psi}{:}E\ket{\psi}).
\end{align*}\qed

Note that the number of qubits $n$ is simple relative to $E$, thus the complexity terms in the proof are relativized to $n$. An implication of Lemma \ref{lmm:meas}  is that most pure quantum states have almost no self classical information. That is for an overwhelming majority of pure quantum states, the probabilities induced by a measurement will have negligible self information, as shown in Theorem \ref{thr:noinfo}. 

\begin{thr}
	\label{thr:noinfo}
	Let $\Lambda$ be the uniform distribution on the unit sphere of an $n$ qubit space. Relativized to POVM $E$, $\int 2^{\i(E\ket{\psi}:E\ket{\psi})}d\Lambda = O(1)$.
\end{thr}	
\prf 
By Lemma \ref{lmm:meas}, $2^{\I(\ket{\psi}:\ket{\psi})}\gem 2^{\i(E\ket{\psi}:E\ket{\psi})}$. From Lemma \ref{lmm:selfinfo}, $\int 2^{\I(\ket{\psi}:\ket{\psi})}d\Lambda=O(1)$. The integral $\int 2^{\i(E\ket{\psi}:E\ket{\psi})}d\Lambda $ is well defined because $2^{\i(E\ket{\psi}:E\ket{\psi})}=\Tr \sum_{i,j}\nu_{i,j}2^{-\K(i,j)}(\ket{\psi}\bra{\psi}\otimes \ket{\psi}\bra{\psi})$, which can be integrated over $\Lambda$.\qed\\

Theorem \ref{thr:noinfo} says that given a measurement apparatus, the overwhelming majority of pure states, when measured, will produce classical probabilities with no self-information, i.e. random noise. Theorem \ref{the:classconsinfo} shows that there is no randomized way to process the probabilities to produce more self-information, i.e. process the random noise.

\appendix
\section{Conservation of Information Between Probabilities}
\label{sec:classconsinfo}
In this section, we show that the information between probabilities cannot be increased through randomized transformations. We recall that information $\i$ was introduced in Section \ref{sec:measure}, where for probabilities $p$ and $q$, $\i(p:q)=\sum_{x,y}2^{\i(x:y)}p(x)q(y)$. The information between strings $x,y\in \FS$ is $\i(x:y)=\K(x)+\K(y)-\K(x,y)$. A probability $p$ is transformed by channel $f:\FS\times\FS\rightarrow\R_{\geq 0}$, by $fp(x)=\sum_{y}f(x|y)p(y)$. For channel $f$, $f(\cdot|y)$ is a conditional probability given $y\in\FS$.

\begin{lmm}[\cite{Levin84}]For $x,y,z\in\FS$,
	$\i(x:y)\lea \i((x,z):y)$.
	\label{lem:infmon}
\end{lmm}
\begin{lmm}
	\label{lem:consran}
	Let $\psi_d$ be a semi-measure,  lower computable relative to $d\in\FS$. For $a,b\in\FS$\\ 
	$\sum_{c\in\FS}2^{\i((a,c):b)}\psi_a(c)\lem 2^{\i(a:b)}$.
\end{lmm}
\prf
	This requires a slight modification of the proof of Proposition 2 in \cite{Levin84}. We need to show $\m(a,b)/(\m(a)\m(b))\gem \sum_c (\m(a,b,c)/(\m(b)\m(a,c)))\psi_a(c)$, or $\sum_c (\m(a,b,c)/\m(a,c))\m(c|a)\lem \m(a,b)/\m(a)$, since $\m(c|a)\gem\psi_a(c)$. Rewrite it $\sum_c\m(c|a)\m(a,b,c)/\m(a,c)\lem \m(a,b)/\m(a)$ or $\sum_c\m(c|a)\m(a)\m(a,b,c)/\m(a,c)\lem \m(a,b)$. The latter is  true since $\m(c|a)\m(a)\lem \m(a,c)$ and $\sum_c\m(a,b,c)\lem \m(a,b)$.
\qed
\begin{thr}
	\label{the:classconsinfo}
	For probabilities $p$ and $q$, relativized to channel $f$, $\i(fp:q)\lea\i(p:q)$.
\end{thr}
\prf Using Lemma \ref{lem:infmon},
\begin{align*}
	\i(fp:q)&=\log\sum_{x,y}2^{\i(x:y)}\sum_zf(x|z)p(z)q(y)\lea\log\sum_{y,z
	}q(y)p(z)\sum_x2^{\i((z,x):y)}f(x|z).
\end{align*}
Using Lemma \ref{lem:consran},
\begin{align*}
\i(fp:q)&\lea\log\sum_{z,y}q(y)p(z)2^{\i(z:y)} \eqa \i(p:q).
\end{align*}

\newcommand{\etalchar}[1]{$^{#1}$}

\end{document}